\documentclass[setspace,amsmath,amssymb]{article}
\usepackage[all]{xy}
\usepackage{graphicx}
\usepackage{setspace}
\usepackage{amsmath}
\usepackage{amssymb}
\usepackage{rsfs}
 
\newcommand{\ra}{\rangle}

\newcommand{\be}{\begin{equation}}
\newcommand{\ee}{\end{equation}}
\newcommand{\bea}{\begin{eqnarray}}
\newcommand{\eea}{\end{eqnarray}}

\begin{document}
\begin{titlepage}

\begin{flushright}
\today
\end{flushright}

\vspace{1in}

\begin{center}

{\bf More on quantum measuring systems and the holographic principle}

\vspace{1in}

\normalsize

{Eiji Konishi\footnote{E-mail address: konishi.eiji.27c@kyoto-u.jp}}

\normalsize
\vspace{.5in}

{\it Graduate School of Human and Environmental Studies,\\
 Kyoto University, Kyoto 606-8501, Japan}
\end{center}

\vspace{1in}

\baselineskip=24pt
\begin{abstract}
In this article, we approach the structure of the quantum measuring system in the Euclidean regime of the classicalized holographic tensor network from the perspective of integrated information theory.
As a result, we obtain the following picture of the Euclidean regime.
First, there are complexes, which are independently accompanied by the level and structure of experiences, determined from the full transition probability matrix of the whole particle system.
Second, the cause--effect structures of independent complexes would be directly entangled by the physical information propagation in the whole particle system.
Finally, distinct full transition probability matrices of the whole particle system that exhibit the maximum cause--effect power may coexist.
\end{abstract}

\vspace{.1in}


\vspace{.6in}


\end{titlepage}


\section{Introduction}

This article is the third in a series \cite{JHAP4,JHAP6} to investigate the links between quantum measuring systems and the holographic principle \cite{Hol1,Hol2,Hol3,AdSCFT1,AdSCFT2}.
In the first article \cite{JHAP4}, we conclude that the particles are always substantiated to exist when we consider the Euclidean regime of the Universe that is modeled as the classicalized holographic tensor network (cHTN) in three spacetime dimensions \cite{Swingle,MIH,Bao,EPL1,EPL2,JHAP1}, in addition to its Lorentzian regime.
In the second article \cite{JHAP6}, we show the continuity of the subject of quantum measurement through the same consideration of the Euclidean regime of the cHTN.
In this third article, we model and analyze the structure of the quantum measuring system in the Euclidean regime of the cHTN in the framework of integrated information theory (IIT) \cite{IIT3,IIT4}.

In the main part of this article, we first give an overview of the quantum measuring system in the Euclidean regime of the cHTN.
Second, we directly apply the framework of IIT to this regime.
Then, the {\it complex} is defined (see equation (\ref{eq:complex})), and the difference between the quantum measurement schemes in the Lorentzian and Euclidean regimes is incorporated into the analysis.

The rest of this article is organized as follows.
In Section 2, we introduce the terminology and basic ideas in IIT as preliminaries.
In Section 3, we investigate aspects of a quantum measuring system (the human brain) in the Lorentzian regime as a role model, irrespective of the holographic principle.
The quantum measuring system is in our Lorentzian Universe.
Section 4 is the main part of this article.
We consider the holographic principle in the context of the classicalization, and we model and analyze the structure of the quantum measuring system in the Euclidean regime of the cHTN in the framework of IIT.
In Section 5, we conclude this article.

\section{Preliminaries}

In this section, we introduce the terminology and basic ideas of IIT 3.0 and 4.0 \cite{IIT3,IIT4}.

Here, we assume a stochastic (coarse-grained) dynamical system $X$ that is a discrete set of elements.
Then, we consider a subset of system $X$ in the present, and subsets of system $X$ in the past and future by one discrete dynamical step.
In the IIT literature on \cite{IIT3,IIT4}, the former subset is called a {\it mechanism}, and the latter subsets are called {\it purviews}.

Once the full transition probability matrix (TPM) of system $X$ is given in its two states, all of the contents of IIT with respect to system $X$ follow.
First of all, from the full TPM and the uniform marginal distributions, we can construct the cause/effect TPM defined for the pair of a given mechanism and its past/future purview in their states.
The purview need not match with the mechanism as a set, and the state of the mechanism is determined.
From the cause TPM, the core past purview of the mechanism is determined, and from the effect TPM, the core future purview of the mechanism is determined independently.
Here, these {\it core} purviews maximize the irreducibility (i.e., the integration) of the cause--effect power of the mechanism within the dynamical causal structure of system $X$.
In IIT 4.0, the pair of the core cause and the core effect associated with a mechanism is called a {\it distinction}.

In IIT, we quantify the maximum irreducibility of the cause--effect power of a mechanism by considering the possible sets of cuts (here, a {\it cut} is the coarse-graining operation of the cause/effect TPM of an element of the past/future purview with respect to the state of an element of the mechanism) of the cause/effect TPM and calculating the minimum information distance between the two probability distributions of the past/future purview before and after the cuts of the cause/effect TPM.
The minimum between these cause and effect quantities is called {\it integrated information}.
This is primary mechanism-level integrated information, and there is also secondary system-level integrated information.

As the central identifications in IIT, the distinctions and their {\it relations} (overlaps of the purviews of possible distinctions) of system $X$ obtained at the mechanism level are identified with the {\it structure} of an experience of system $X$, and the system-level integrated information of system $X$ is identified with the {\it level} of the experience of system $X$.

\section{Lorentzian regime: the role model}

In this section, we first identify the four systems in the two groups
\begin{equation}
(S_0,A^\prime)_S\;,\ \ (\psi,A)_M
\end{equation}
that appear in the abstract framework of the projective quantum measurement in the ensemble interpretation of quantum mechanics \cite{dEspagnat,JSTAT1,EPL3,IJQI} with concrete systems in the human brain.\footnote{For the details of this abstract framework, see Refs. \cite{EPL3,IJQI}.}
This identification is based on the author's recent work \cite{QSMF,arXiv}.
This section presents the role model of the model in the next section.

A combined measured system $S:=S_0+A^\prime$ performs a {\it non-selective measurement}, which is the dynamical quantum measurement process from a given quantum pure state (i.e., a quantum superposition of state vectors of the measured system $S_0$ in the eigenbasis of the discrete measured observable) to its classical mixed state (i.e., the diagonal quantum mixed state of eigenstates of the measured observable).

The measured system $S_0$ is identified with the system of an external stimulus (sound, light, force, heat, etc.) inputted into the brain: \cite{QSMF}
\begin{equation}
S_0={\rm External\ stimulus}\;.
\end{equation}

The macroscopic measurement apparatus $A^\prime$ that performs the non-selective measurement by its von Neumann-type interaction with system $S_0$ is identified with the macroscopic sensory organ $A_{\rm so}$ in question: \cite{QSMF}
\begin{equation}
A^\prime=A_{\rm so}\;.
\end{equation}

Throughout this section, {\it macroscopicity} of a quantum mechanical system refers to the applicability of the superselection rule (i.e., the classicality) to the canonical variables of the quantum mechanical system \cite{dEspagnat,EPL3,IJQI}.

The measuring system $M:=\psi+A$ performs an {\it event reading}, which is the informatical quantum measurement process from the classical mixed state, obtained by the quantum entanglement process between systems $S_0$ and $\psi$ after the non-selective measurement of system $S$, to a classical pure state (i.e., a simultaneous eigenstate of the measured observable and the following discrete meter variables of system $\psi$).

The quantum mechanical event-reading system $\psi$ is identified with the neurons in a complex of the brain, each of which is in the neural rest or firing state \cite{arXiv}.
The discrete meter variables $\{{\mathfrak M}\}$ of system $\psi$ are identified with the binary observables of system $\psi$, each of which distinguishes the rest and firing states of a neuron by ${\mathfrak M}=0$ and ${\mathfrak M}=1$, respectively \cite{arXiv}.
A fine-grained read-out state of system $\psi$ at a given real time $t$ is
\begin{equation}
\psi_t=\{{\mathfrak M}_t\}\;,\ \ {\mathfrak M}_t=0\ \ {\rm or}\ \ 1\;.
\end{equation}

A quantum-field-theoretical macroscopic Bose--Einstein condensate $A$ that induces a trigger of an event reading by the von Neumann-type interaction (here, the Coulomb interaction) of its center of mass with system $\psi$ \cite{EPL3} is identified with that of evanescent (tunneling) photons $A_{e^\ast, m^\ast}$ created in the perimembranous region of an individual neuron by the Anderson--Higgs mechanism: \cite{arXiv,JPY,JY}
\begin{equation}
A=A_{e^\ast, m^\ast}\;.
\end{equation}
Here, the evanescent photons are not only massive but also electrically charged \cite{JY}.
$e^\ast$ and $m^\ast$ are the non-vanishing effective electric charge and non-vanishing effective mass of an evanescent photon, respectively, and, for the Anderson--Higgs mechanism, the Goldstone boson is the {\it electric dipole phonon} in the water molecules (electric dipoles with macroscopic order) in the perimembranous region of the neuron \cite{JY}.

According to IIT \cite{IIT3,IIT4}, the level and structure of an {\it experience} (i.e., the coarse-grained read-out state with the maximum cause--effect power; see footnote 3) $\psi_t$ of system $\psi$ at a given real time $t$ are determined by the cause and effect TPMs, which are
\begin{eqnarray}
{\cal T}_c&=&p_c(\psi^{\prime\prime}_{t-1}|\psi^\prime_t)\;,\\
{\cal T}_e&=&p_e(\psi^{\prime\prime\prime}_{t+1}|\psi^\prime_t)\;,
\end{eqnarray}
respectively.
Here, in the cause (resp., effect) conditional probability function $p_c$ (resp., $p_e$), the {\it present} experience $\psi^\prime_t$ of a mechanism $\psi^\prime$ is determined and, to analyze the composition structure of the experience $\psi_t$, we consider all of the subsets of $\psi$:
\begin{equation}
\psi^\prime\ ({\rm mechanism})\;,\ \ \psi^{\prime\prime}\;,\ \psi^{\prime\prime\prime}\ ({\rm purviews})\ \subseteq \psi\ ({\rm complex})\;.
\end{equation}

Whether or not physicality accompanies an experience $\psi_t$ of system $\psi$ is determined by the criterion in terms of the work required for event reading $W_{\rm e.r.}$:
\begin{equation}
W_{\rm e.r.}=k_BT\ \ {\rm or}\ \ 0\;,
\end{equation}
where $T$ refers to the constant temperature of the thermal environment of system $S$ \cite{JSTAT1,IJQI}.
$W_{\rm e.r.}$ is {\it classical} work \cite{IJQI} and is done by system $M$ on system $S$ \cite{JSTAT1,IJQI}.

Here, when the non-selective measurement is done via a sensory organ (i.e., $S\neq M$), $W_{\rm e.r.}=k_BT$ holds and physicality accompanies the experience $\psi_t$; otherwise (i.e., $S=M$), $W_{\rm e.r.}=0$ holds and physicality does not accompany the experience $\psi_t$ \cite{JSTAT1,IJQI}.

\section{Euclidean regime}

In this section, we study the quantum measuring system in the Euclidean regime of the three-dimensional cHTN spacetime.
This section is the main part of this article.

\subsection{Overview}

In the arguments of the holographic principle, we assume that the quantum state of the hologram ${\cal H}$ on the boundary spacetime is classicalized \cite{EPL1,EPL2}.
This quantum state is the ground state $|\psi\ra_{\rm CFT}$ of a strongly coupled conformal field theory (CFT) on the boundary spacetime with an Abelian restricted set ${\cal A}$ of the qubits observables in the boundary CFT:
\begin{equation}
{\cal H}=(|\psi\ra_{\rm CFT},{\cal A})
\end{equation}
in the presence of the superselection rule operator $\sigma_3$ (the one-qubit third Pauli matrix) \cite{JHAP1}.

In bits, the von Neumann entropy $S_{\rm v.N.}$ of the classical mixed state, with respect to $\sigma_3$, of the classicalized hologram ${\cal H}$ is the discretized area $A_{\rm TN}$ of the cHTN space: \cite{EPL2}
\begin{equation}
S_{\rm v.N.}[{\cal H}]=A_{\rm TN}\ \ {\rm bits}\;.
\end{equation}

Now, we consider non-relativistic particles in the absence or presence of their interactions in the bulk spacetime.
We denote the full set of the read-out imaginary-time trajectories $\gamma_\tau$ of the particles in the bulk spacetime by
\begin{equation}
\Gamma_\tau=\{\gamma_\tau\}\;.
\end{equation}
The increment of $S_E^{(0)}$, which is the Euclidean action of the particles minus the potential energy part, per $\hbar b$ ($b=\ln 2$) gives rise to a locally defined {\it spin}-event reading from the cHTN \cite{JHAP1}.
The total amount of the acquired information $I$ with respect to the spin events is
\begin{equation}
I=\frac{S_E^{(0)}[\Gamma_\tau]}{\hbar b}\ \ {\rm bits}\;.
\end{equation}
Note that the spin-event reading occurs at a particular site (i.e., the top tensor of the time slice of a particular entanglement wedge) of the cHTN, and then the positional information of the imaginary-time trajectories $\gamma_\tau$ is also read by the classicalized hologram ${\cal H}$.

In the IIT literature on \cite{IIT3,IIT4}, a subset $\Gamma^\ast$ of a given physical substrate $U$ (the present state $U_\tau$ of $U$ and the full TPM of $U$ are also given) is called a {\it complex} if it satisfies the criterion
\begin{equation}
\Gamma^\ast=\underset{\Gamma\subseteq U}{\operatorname{argmax}}\varphi_\Gamma (\Gamma_\tau)\;,\ \ \Gamma_\tau=U_\tau|_\Gamma\label{eq:complex}
\end{equation}
for the system integrated information $\varphi_\Gamma$ (for its mathematical construction from the present state $U_\tau$ of $U$ and the full TPM of $U$, see Ref. \cite{IIT4}).\footnote{In the complex, the grain is chosen so that the system integrated information will be maximized \cite{IIT3,IIT4}.}

In particular, this complex satisfies the fourth axiom of IIT 4.0 \cite{IIT4}, that is, the {\it exclusion} property (i.e., the definiteness) of its experience due to the maximization criterion of the system integrated information.

According to IIT \cite{IIT3,IIT4}, the level and structure of an experience $\Gamma^\ast_\tau$ of complex $\Gamma^\ast$ at a given imaginary time $\tau$ are determined by the cause and effect TPMs, which are
\begin{eqnarray}
{\cal T}_c&=&p_c(\Gamma^{\ast \prime \prime}_{\tau - 1}|\Gamma^{\ast \prime}_\tau)\;,\\
{\cal T}_e&=&p_e(\Gamma^{\ast \prime \prime\prime}_{\tau + 1}|\Gamma^{\ast \prime}_\tau)\;,
\end{eqnarray}
respectively.
Here, the coarse-grained spatial positions in the read-out imaginary-time trajectories at a given imaginary time are the events in the conditional probability functions $p_c$ and $p_e$ and, to analyze the composition structure of the experience $\Gamma^\ast_\tau$, we consider all of the subsets of $\Gamma^\ast$:
\begin{equation}
\Gamma^{\ast \prime}\ ({\rm mechanism})\;,\ \ \Gamma^{\ast \prime\prime}\;,\ \Gamma^{\ast \prime \prime\prime}\ ({\rm purviews})\ \subseteq \Gamma^\ast\ ({\rm complex})\;.
\end{equation}

Finally, physicality does not accompany any experience $\Gamma^\ast_\tau$ of complex $\Gamma^\ast$ because
\begin{equation}
W_{\rm e.r.}\ ({\rm done\ by}\ {\cal H})=0\label{eq:zero}
\end{equation}
always holds in the Euclidean regime.
Equation (\ref{eq:zero}) is zero because there is no non-selective measurement process in the Euclidean regime of the cHTN.

\subsection{Entanglement process}

In the Euclidean regime, we assume the full TPMs
\begin{eqnarray}
{\cal T}_1&=&p(\Gamma^\ast_{1,\tau+1}|\Gamma^\ast_{1,\tau})\;,\\
{\cal T}_2&=&p(\Gamma^\ast_{2,\tau+1}|\Gamma^\ast_{2,\tau})
\end{eqnarray}
of two independent complexes $\Gamma^\ast_1$ and $\Gamma^\ast_2$ of the whole particle system, respectively.
Then, we consider the entanglement process of these full TPMs:
\begin{equation}
{\cal T}_1\otimes {\cal T}_2\to {\cal T}_{1,2}\;,
\end{equation}
where ${\cal T}_{1,2}$ is entangled with respect to $\Gamma^\ast_1$ and $\Gamma^\ast_2$.

In the Lorentzian regime, such a process requires the Shannon communication protocol because the projective quantum measurement requires the non-selective measurement process before event reading.

However, in the Euclidean regime, this process is directly done by physical information propagation in a larger particle system $\Gamma$, which contains both complexes $\Gamma^\ast_1$ and $\Gamma^\ast_2$.
Note that, in the Euclidean regime, the boundary of each complex is formed by the system integrated information only.

From this argument, we obtain an intuitive picture of the Euclidean regime:
there are complexes, which are independently accompanied by the level and structure of experiences, determined from the full TPM of the whole particle system, and the cause--effect structures of independent complexes would be directly entangled by the physical information propagation in the whole particle system.

\section{Conclusion}

In this article, after the preliminaries of IIT, first we investigated aspects of a quantum measuring system (the human brain) in the Lorentzian regime of our Universe.
Next, based on the classicalization of the holographic tensor network in three spacetime dimensions, we modeled and analyzed the structure of the quantum measuring system in the Euclidean regime of the cHTN in the framework of IIT.

We conclude by providing a novel picture of the quantum measuring system in the Euclidean regime of the cHTN.

First, in the Euclidean regime, the contents of IIT are determined by the full TPM
\begin{equation}
{\cal T}_{\cal U}=p({\cal U}_{\tau+1}|{\cal U}_\tau)\;,
\end{equation}
of the whole particle system ${\cal U}$ in the cHTN, that exhibits the maximum cause--effect power.
Of course, the coarse-graining operations in system ${\cal U}$ are incorporated into the definition of this TPM.
This TPM can be approximately split into $N$ tensor product of the full TPMs:
\begin{equation}
{\cal T}_{\cal U}\sim \left(\bigotimes_{i=1}^N{\cal T}_i\right)\;.
\end{equation}
Each of the full TPMs ${\cal T}_i$ ($i=1,2,\ldots,N$) defines a complex $\Gamma^\ast_i$ of system ${\cal U}$.\footnote{This criterion on the complex is stronger than the definition (\ref{eq:complex}) of the complex.}
An experience $\Gamma^\ast_{i,\tau}$ of each complex $\Gamma^\ast_i$ is exclusive (i.e., definite) \cite{IIT3,IIT4}, and the level and structure of the experience $\Gamma^\ast_{i,\tau}$ are independent from the other complexes.
The boundary of each complex is formed by only the system integrated information via the criterion (\ref{eq:complex}).

Second, in the Euclidean regime, to entangle the full TPMs of independent complexes, physical information propagation in a larger particle system that contains the complexes is required, and the Shannon communication protocol used in the Lorentzian regime is not necessary at all.
This is because the quantum measurement in the Euclidean regime does not require the non-selective measurement process before event reading.

Finally, for system ${\cal U}$, the definition of the full TPM ${\cal T}_{\cal U}$ that exhibits the maximum cause--effect power may not be unique.
Indeed, there are many essential coarse-graining factors in system ${\cal U}$
(the spatiotemporal grain, and the grain of elements, states, and updates in system ${\cal U}$ \cite{IIT4})
that may lead to distinguishing the full TPMs, of system ${\cal U}$, that exhibit the maximum cause--effect power
\begin{equation}
{\cal T}_{{\cal U},1}\;,\ {\cal T}_{{\cal U},2}\;,\ldots\;.\label{eq:TPM}
\end{equation}
From each of these, the distinct set of complexes is constructed; each of these full TPMs describes the same system ${\cal U}$ and coexists with the others in this sense.
However, at the present level of analysis, the coexistence of many full TPMs (\ref{eq:TPM}) that exhibit the maximum cause--effect power is nothing more than a prospect.

\end{document}